\documentclass{article}
\usepackage{spconf,amsmath,graphicx}
\usepackage{algorithmicx}
\usepackage{algorithm}
\usepackage{algpascal}
\usepackage{algc}
\usepackage{algcompatible}
\usepackage[noend]{algpseudocode}
\usepackage{caption}
\usepackage{subcaption}
\usepackage{float}
\usepackage[nounderscore]{syntax}
\usepackage[utf8]{inputenc}
\usepackage{amsmath,graphicx}

\newcommand{\algrule}[1][.2pt]{\par\vskip.5\baselineskip\hrule height #1\par\vskip.5\baselineskip}

\DeclareMathOperator*{\NPoints}{NPoints}
\DeclareMathOperator*{\QVal}{QVal}

\title{Optimizing human-interpretable dialog management policy using Genetic Algorithm}
\name{Hang Ren, Weiqun Xu and Yonghong Yan
}
\address{The Key Laboratory of Speech Acoustics and Content Understanding \\
  Institute of Acoustics, Chinese Academy of Sciences\\
  21 North 4th Ring West Road, Beijing, China, 100190}
\begin{document}
%
\maketitle
\begin{abstract}
	Automatic optimization of spoken dialog management policies that are
	robust to environmental noise has long been the goal for
	both academia and industry.
	Approaches based on reinforcement learning have been proved to be effective.
	However, the numerical representation of dialog policy is human-incomprehensible
	and difficult for dialog system designers to verify or modify,
	which limits its practical application.
	In this paper we propose a novel framework for optimizing
	dialog policies specified in domain language using genetic algorithm.
	The human-interpretable representation of policy makes the method suitable for practical employment.
	We present learning algorithms using user simulation and real human-machine dialogs respectively.
	Empirical experimental results are given to show the effectiveness of the proposed approach.
\end{abstract}
\section{Introduction}
\label{sec:intro}

Dialog manager (DM) plays a central part in spoken dialog system (SDS) and
its major functionalities include tracking dialog states and
maintaining a dialog policy which decides how the system reacts given certain dialog state.
Designing a dialog policy by hand is tedious and erroneous because of the uncertainty of underlying dialog states
especially in noisy environment.
In recent years various approaches for automatic DM policy optimization have been proposed \cite{williams_partially_2007,young_hidden_2010,lee_example-based_2009,lison_structured_2014},
among which methods based on reinforcement learning (RL) and POMDP model are the most popular \cite{young_pomdp-based_2013}.
The main objective of RL is to learn an optimum policy conducted by an agent
by maximizing its cumulative reward.
One of the advantages of RL-based DMs is its robustness to noises from automatic speech recognizer (ASR) 
and spoken language understanding (SLU) modules.
Also, it automates the optimization process by allowing the agent to discover the optimum policy
through exploring the underlying state-action space and incrementally improve the controlling policy.

Despite all the advantages, RL-based DMs are not widely deployed for commercial SDSs
due to several reasons 
\cite{paek_reinforcement_2006}.
Firstly, RL algorithms are mostly data-demanding,
which leaves dialog system designers in a dilemma since there is usually few or even no data available
at the early stage of system development.
Several methods have been proposed to mitigate this problem.
A user simulator is often firstly built using wizard-of-oz dialog data,
and then the simulator is used to train a RL-based DM.
In recent studies it has been shown that by incorporating domain knowledge into the design of kernel functions,
the GPSARSA \cite{engel_reinforcement_2005,gasic_gaussian_2010} algorithm
exhibits much faster learning speed than conventional online RL methods.
Secondly, RL algorithms usually use complex numerical models in optimizing the value function,
which are usually beyond human comprehension.
The learned policy is implicitly represented in the optimized value function (Q-function),
which is difficult or even impossible for system designer to verify or modify,
keeping back domain experts from
setting necessary constraints over the system behavior.

In this paper we propose to use Genetic Algorithm (GA) \cite{whitley_genetic_1994} in optimizing DM policies (GA-DM)
which are comprehensible to human designers and easy to verify and modify.
The underlying idea is intuitive.
We use human-readable domain language to sketch the basic structure of the DM policy,
and leave the uncertain parameters for later tuning.
According to our experiences in deploying SDSs,
it is relatively easy to specify a basic DM policy,
when engineering slot-filling or task-driven SDSs of a moderate scale.
The most difficult part lies in setting various threshold parameters 
in dealing with ASR and SLU errors
via repeatedly confirming and grounding.
These parameters are usually set heuristically or by trail-and-error.
Automatic optimization of these parameters will be of great help.
We hope to keep the trade-off between purely hand-designed rule-based policies
and the ones automatically learned using black-box and data-driven RL methods
while keeping the merits from both approaches.
Two variants of the approach are proposed and evaluated, 
an on-line training method through interaction with a simulated user and
an off-line and sample-efficient version called on-corpus Q-points regression.

In the following sections we describe the algorithms and experiments in detail.
In section \ref{ga_dpt} we briefly describe Genetic Algorithm and its application in DM policy optimization.
We propose two different policy optimization methods based on simulation and dialog corpus
in sections \ref{simulation} and \ref{on_corpus} respectively.
In section \ref{experiments} we give experimental results on simulated user and real human-machine dialog corpus.

\begin{figure}[t]
\centering
\includegraphics[scale=0.65]{{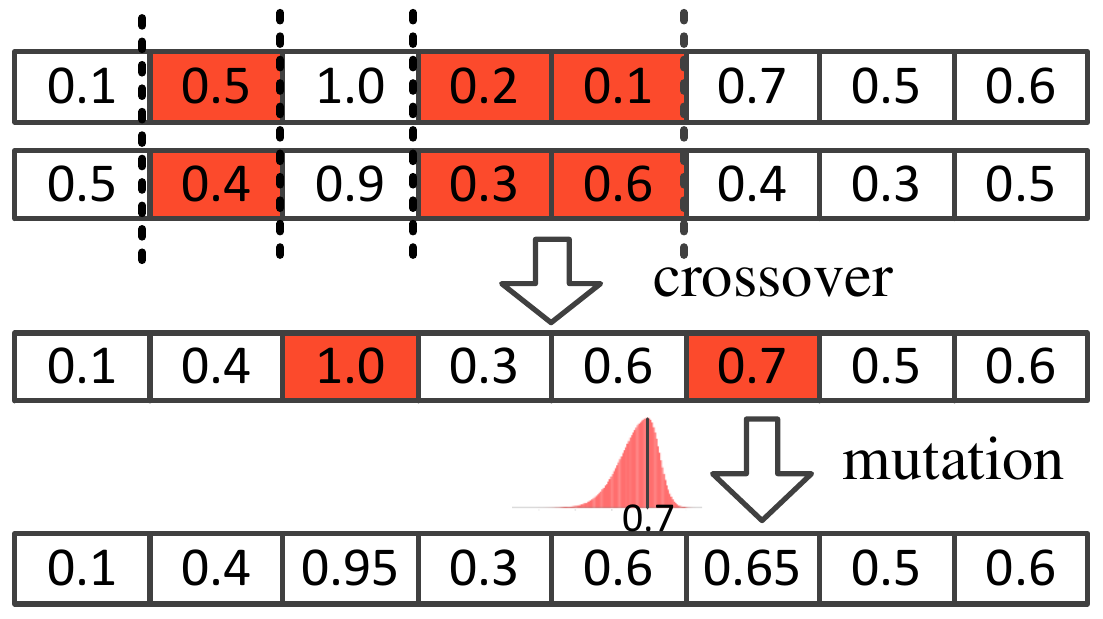}}
\caption{Crossover and mutation of individuals (chromosomes). 
Each individual is a real vector with consitituent scalars in $[0, 1]$.
}
\label{crossover_mutation}
\end{figure}

\newfloat{algorithm}{th!}{lop}
\begin{algorithm}
\caption{Genetic algorithm policy optimization\label{ga}}
\begin{algorithmic}[1]
 \State \textbf{Input} fitness function $F$,
	 $N_{pop}$, $N_{mut}$, $T_{max}, K$, $\sigma$, $\mu_{mut}$
 \State $t \gets 0, P_0 \gets \emptyset$ \Comment{the initial population}
 \For{$i \gets 1, \ldots, N_{pop}$ }
	 \State $P_0$.\textit{add}(\textit{Random.generateIndividual}()) \Comment{random initialization}
 \EndFor
 \State $P_0$.\textit{evalFitness}() \Comment{evaluate fitness of each individual}
 \While{fitness $f_t$ not converges \textbf{and} $t<T_{max}$}
	 \State $t \gets t + 1$,  $P_t \gets \emptyset$\Comment{next generation}
	 \State $P_t$.\textit{add}($P_{t-1}$.\textit{getFittest}())\Comment{elitism}
	 \For{$i \gets 1, \ldots, N_{mut}$ }
		 \State $P_t$.\textit{add}(\textit{mutate}($P_{t-1}$.\textit{getFittest}(), $\sigma$, $\mu_{mut}$)) \Comment{mutate the fittest}
	 \EndFor
	 \For{$i \gets 1, \ldots, N_{pop}-N_{mut}-1$ }
		 \State $I_1, I_2 \gets$ \textit{tournamentSelect}$(P_{t-1}, K)$
		 \State $P_t$.\textit{add}(\textit{mutate}(\textit{crossover}($I_1, I_2$), $\sigma$, $\mu_{mut}$)) \Comment{reproduction}
	 \EndFor
	 \State $P_t$.\textit{evalFitness}()
	 \State $f_t=P_t$.\textit{getFittest}().\textit{getFitness}()
 \EndWhile
\State \Return $P_t.\textit{getFittest}()$
\algrule
\Function{mutate}{$I$, $\sigma$, $\mu_{mut}$} \Comment{mutate an individual $I$}
\For{each parameter $\theta_i$ of $I$}
	\If{\textit{Random.uniform}() $< \mu_{mut}$}
		\State $I.\theta_i \gets \textit{perturb}(I.\theta_i, \sigma)$
	\EndIf
\EndFor
\State \Return $I$
\EndFunction
\algrule
\Function{perturb}{$\theta$, $\sigma$} \Comment{add random noise to a single parameter}
	\State $g \gets$ \textit{abs}(\textit{Random.stdGaussian}())
	\If{\textit{Random.uniform}() $< \theta$}
		\State $v \gets -\frac{g}{\sigma}*\theta + \theta$
	\Else
		\State $v \gets \frac{g}{\sigma}*(1.0-\theta) + \theta$
	\EndIf
	\If{$v < 0.0$ \textbf{or} $v > 1.0$}
		\State \Return \textit{perturb}($\theta$, $\sigma$)
	\Else
		\State \Return $v$
	\EndIf
\EndFunction
\algrule
\Function{tournamentSelect}{$P$, $K$} \Comment{tournament selection}
	\State choose a random subset $P_K$ of size $K$ from $P$
	\State \Return $P_K$.\textit{getFittest}()
\EndFunction
\algrule
\Function{crossover}{$I_1, I_2$} \Comment{crossover of two parents}
	\State $I' \gets$ exchange random parts of $I_1$ and $I_2$
	\State \Return $I'$
\EndFunction
\end{algorithmic}
\end{algorithm}

\section{Models and Algorithms}
\label{model_algo}

\subsection{Genetic algorithm and dialog policy template}
\label{ga_dpt}

\newfloat{Listing}{t}{lop}
\begin{Listing}
\begin{grammar}
<template> ::=
`if' <cond-exp> `then' <action> `else' <template>
\alt `if' <cond-exp> `then' <action> `else' <action>

<cond-exp> ::= <cond-exp> <logic-op> <cond-exp>
\alt <boolean-state-var>
\alt <num-state-var> <comparator> <free-param>

<comparator>  ::= `<' | `>' | `=='

<logic-op> ::= `and' | `or'
\end{grammar}
\caption{BNF grammar of dialog policy template}
\end{Listing}

Genetic algorithm is a general optimization framework.
It simulates the evolution process of natural selection by
keeping a \textit{population} of candidate solutions (individuals) and incrementally improve the quality 
using various genetic operators.
It is a \textit{global} optimization method
which can solve both numerical and combinatorial problems.
The key constituent of GA is a \textit{fitness} function evaluating the utility
of each individual.
GA has been proved to be effective
 in solving various problems,
including optimizing controllers in AI games.
The psudocode of optimizing DM using GA is given in Algorithm \ref{ga}.
We refer readers to \cite{whitley_genetic_1994} for a detailed description of GA.
The concepts of \textit{genotype} and \textit{phenotype} are not discriminated here.
In GA an individual directly carries all the information comprising a solution,
which is a fixed-length floating-point array in our experiment and
each number is in $[0, 1]$ as a free parameter of the dialog policy template.
An individual can instantiate a concrete DM policy,
with a defined policy template.
The policy template is composed of a set of prioritized condition-action expressions
and used to specify the basic structure of a dialog policy.
Given certain dialog state,
each condition expression is checked sequentially and
the first matched one is selected with 
the associated action chosen as output.
Listing 1 gives the BNF grammar of the proposed templates.
The actions of the template is fixed and free parameters can be used
to set thresholds for numerical state variables.
Apart from the conditional expression,
parameters can also be used to induce new state variables,
for example a variable representing the number of slots 
whose top scores are above certain threshold.
Although the general system action is fixed in the template,
the `structure' of the action 
(in this slot-filling setting, 
structure includes sub-dialog-actions and the associated slots and values)
can be controlled by parameters.
For example, in the action `offer',
threshold can be used to filter the hypotheses that are
used in searching for the queried information.

Note that the template in Listing 1 has been proposed for its
conciseness and simplicity and does not have to be fixed.
The design of the dialog template 
requires knowledge in the dialog domain
but does not need a exact model of the environmental noise,
thus is very suitable for human experts.
This engineering division is intentionally made in our proposed approach.

In GA two kinds of genetic operators are used, i.e.
\textit{mutation} and \textit{crossover},
which are shown graphically in Figure \ref{crossover_mutation}
and as pseudo-code in Algorithm \ref{ga}.
During crossover, two parents are selected,
then random parts of the two parents are exchanged,
giving birth to a new child.
The mutation operator checks each component of a chromosome sequentially,
either leaving it intact or perturbing it randomly.
In our implementation the perturbation is realized
by sampling from a skewed normal distribution
with the peak centered at the perturbed number.
If the sampling result lies outside $[0, 1]$,
the process is repeated by calling the function \textit{perturb} recursively.
This sampling sub-routine is designed for 
a smooth distribution function.
The \textit{mutation} and \textit{crossover} operators represent
asexual and bisexual reproductions in GA respectively.
Other reproducing strategy can be used as long as
it effectively explores the underlying solution space.
\textit{Tournament selection} is used to select individuals for reproduction. 
It is a simple selection method
where random $K$ individuals are chosen from the population.
We also use the \textit{elitism} technique 
passing the fittest individual 
directly to the next generation,
ensuring that the fitness of the population will never decrease.
The fitness function is the most important part of GA since it
 guides the algorithm in searching for optimum solution.
Two kinds of DM policy fitness evaluation methods are described in the following sections.

\subsection{Policy optimization with a user simulator}
\label{simulation}
Since the fitness function should be consistent with
the performance of the DM, 
the most straightforward way is to evaluate it online with users.
But interacting with real user is time-consuming and labor-intensive, 
thus an agenda-based user simulator is utilized \cite{schatzmann_agenda-based_2007} and 
$N$ interactions are conducted between the simulated user
and DM.
Average cummulative reward is used as the fitness for the individual,
which is similar to the objective of common RL algorithms.
\begin{equation}
F_{R}[\pi_{GA}] = \frac{1}{N}\sum_{i=1}^{N} \sum_{j=1}^{l_i} \gamma^{j-1} r_{ij}
\end{equation}
where $r_{ij}$ is the immediate reward and
$\gamma$ the discounted coefficient.

A noisy channel is designed to simulate ASR and SLU errors.
For each dialog act \texttt{\{act, (slot, value)\}},
replacement and deletion are randomly applied to \texttt{value}
given the assigned confidence scores, which are randomly generated too.
The produced N-best hypotheses are then fed into DMs.

\subsection{On-corpus Q-points regression}
\label{on_corpus}

\newfloat{algorithm}{t}{lop}
\begin{algorithm}
\caption{Episodic fitted Q-iteration\label{fitted_q}}
\begin{algorithmic}[1]
 \State \textbf{Input} $\{(s_{i, t}, a_{i, t+1}, s_{i, t+1})\}$ where $t\gets1,\ldots,T_t-1$, and $i \gets 1,\ldots,N$
 \State initialize Q-function approximator $\hat{Q}(s,a)$ and array $Q_{i, t}$ to $0$
 \For{$l \gets 1, \ldots, L_{max}$ }
 	\For{$i \gets 1, \ldots, N$}\Comment{for each dialog}
	 	\For{$t \gets 1, \ldots, T_i-1$}\Comment{for each turn}
		 	 \State $r \gets reward(s_{i, t}, a_{i, t+1}, s_{i, t+1})$
		 	 \If{$t == T_i-1$}
	 	 	 \State $Q_{i, t} \gets r$\Comment{when the dialog ends}
		 	 \Else
	 	 	 \State $Q_{i, t} \gets r + \gamma max_a \hat{Q}(s_{i, t+1}, a)$
		 	 \EndIf
 	    \EndFor
 	\EndFor
 	\State Regress $\hat{Q}(s,a)$ on $\{(s_{i, t}, a_{i, t+1}, Q_{i, t})\}$
 \EndFor
\State Output: $\hat{Q}(s,a)$

\end{algorithmic}
\end{algorithm}

Building a user simulator is not trivial
and it is difficult to measure the consistency of the simulated user behavior to the real one.
Learning a DM using a dialog corpus is appealing but
there is very limited prior work on this subject \cite{henderson_hybrid_2008,Pietquin2011}.
We propose to use an existing dialog corpus to estimate the fitness of a DM.
First, an offline batch RL algorithm is applied on the corpus,
inducing an optimum Q-function $\hat{Q}(s, a)$,
and an implicitly defined policy $\pi_Q(s) = \arg\max_a \hat{Q}(s, a)$
which is optimum with respect to the corpus.
Then $\hat{Q}(s, a)$ is used to define the fitness function.
We use fitted Q-iteration
\cite{ernst_tree-based_2005} to learn a nonparametric
approximator $\hat{Q}(s, a)$, as described in Algorithm 2.
The algorithm uses Bellman equation (line 10)
to update the estimated Q-values.
Extremely Random Trees (ExtraTrees) \cite{geurts_extremely_2006} are utilized
for non-parametric regression.
ExtraTrees are a powerful model for regression and classification
as they are both flexible and less susceptible to over-fitting.
The annotated dialog corpus is represented as
state-action-state triplets in the form of $\{(s_{t-1}, a_t, s_t)\}$,
and used as the training set.
Two fitness estimation methods are proposed based on different heuristics.
For an individual $\pi_{GA}$ whose fitness to be evaluated,
the $\NPoints$ fitness function is used to calculates the number of triplets
where the actions predicted by $\pi_{GA}$ and $\pi_Q$ are identical.
\begin{equation}
F_{\NPoints}[\pi_{GA}] = \sum_i \delta(\pi_{GA}(s_i), \pi_Q(s_i))
\end{equation}
The QVal fitness attempts to estimate the sum of the Q-values 
for the actions predicted by $\pi_{GA}$ on the training triplets.
However, the Q-function trained on a fixed corpus is often 
inaccurate in unexplored regions of the state space \cite{lee_extrinsic_2014,henderson_hybrid_2008}.
To mitigate the problem a supervised classifier $\hat{P}(a|s)$ is built
on the training set with the observed actions as targets.
If the probability for an action is greater than a predefined threshold $\delta$,
the value produced by $\hat{Q}(s, a)$ is used, 
otherwise a constant $R$ is used for punishment.

\begin{align}
F_{\QVal}[\pi_{GA}] =& \sum_i \tilde{Q}_\delta(s_i, \pi_{GA}(s_i)) \\
\tilde{Q}_\delta(s, a) =& 
\begin{cases} 
\hat{Q}(s, a) &\mbox{if } \hat{P}(a|s)>\delta \\\nonumber
R & \mbox{otherwise } 
\end{cases} 
\end{align}
The two fitness functions are different in weighing the importance
of training instances. 
$F_{\QVal}$ will put a greater effort in optimizing
instances with larger potential Q-value 
improvement while avoiding taking unobserved actions.
Combining GA with the above two fitness functions leads to
the on-corpus Q-points regression algorithm.
One limitation of this algorithm compared to the on-line version is that
no free parameter can be present in specifying the action structures
since the fitness functions reply on the result of reinforcement learning,
which does not support dynamical change of action structure.

\subsection{On-corpus DM evaluation}
\label{on_corpus_eval}
We describe a DM evaluation method on dialog corpus
without the need for deploying the DM online.
A held-out dialog corpus is used as testing set,
and the estimated cumulative reward for the testing dialogs
when following the target DM policy
is used as metric for performance.
A similar approach has been taken
in evaluating the effect of different dialog state tracker
on end-to-end performance of a DM \cite{lee_extrinsic_2014}.
The estimation of Q-function is similar to Algorithm \ref{fitted_q}.
But rather than learning the optimum policy,
the value function for the policy to be evaluated is estimated,
with the Bellman iteration (line 10) in Algorithm \ref{fitted_q} changed to:
\begin{equation}
Q_{i, t} \gets r + \gamma \hat{Q}(s_{i, t+1}, \pi(s_{i, t+1}))
\end{equation}
where $\pi$ is the DM policy to evaluate.
Then the average reward for
starting turns $\frac{1}{N}\sum_i Q_{i, 0}$
is used as a metric for performance.

\section{Experiments}
\label{experiments}
We devise a restaurant information domain for dialog simulation.
There are 4 slots for the user simulator to fill before a database query.
During simulations the DM interacts with the user simulator
with a noise channel in between.
The noise level of the channel can be adjusted to simulate different environmental noise conditions.
Since the simulation process is stochastic,
each experiment is conducted for 100 times,
and the mean and standard deviation of testing performance are reported.

The reward function for the simulated environment is defined as follows.
At each dialog turn the agent receives -1.0 reward.
If correct restaurants are offered to users, 100.0 points are rewarded.
But if the information is duplicate to that previously offered or the presented restaurants do not match user goal,
-5.0 points are given.
The reward discounting rate $\gamma$ is set to 0.9.

In the on-corpus evaluation,
DSTC2 dataset \cite{henderson_second_2014} is used for both DM policy learning and evaluation.
The DSTC2 dataset was originally designed as a
benchmark corpus for 
dialog state tracking.
With the detailed annotation of dialog states, actions, SLU outputs and
other information, it can be used as test set for end-to-end DM performance \cite{lee_extrinsic_2014}.
The dialog states used in both simulated and on-corpus experiments
mainly comprise confident scores for each slot.

\subsection{On-line learning experiment by simulation}
\label{simulation_res}
The dialog policy template used in the simulated experiments is shown as follows.
\begin{description}
\setlength\itemsep{0.2em}
\item[c0] On dialog beginning: \texttt{Welcome}
\item[c1] There are no valid SLU results or the top SLU hypothesis score is less than $\theta_0$: \texttt{Repeat}
\item[c2] User has just denied a slot: \texttt{Request} that slot
\item[c3] There is a slot with score less than $\theta_1$ in the tracker: \\\textit{if} the score is larger than $\theta_2$ \textit{then} \texttt{ExplicitConf} \textit{else} \texttt{Request}
\item[c4] The system has not yet output the action \texttt{RequireMore}: \texttt{RequireMore}
\item[c5] Otherwise: query the database with slot-value pairs whose scores are greater than $\theta_3$
\end{description}
The template has 6 condition-action clauses and contains 4 free parameters.
Note that 3 free parameters lie in the condition expressions
while $\theta_3$ is used to adapt the semantics of the macro action \texttt{offer},
which queries the database and presents results to the user.

A rule-based DM policy is built by setting the 4 parameters heuristically.
A RL-based policy trained using Q-learning with linear approximation is also
built for comparison with the proposed methods.
It is trained for over 100,000 dialog sessions 
to ensure that the state-action space is sufficiently explored
and the optimal performance is reached.
The probability for exploration is set to 0.3.
In the training process of each kind of DM, 
the noise level is randomly set for a dialog session.
The same reward function and discounting rate are also used in the fitness estimation of GA.
We run GA for 30 generations in policy training.

\begin{figure}[t]
\centering
\includegraphics[scale=0.65]{{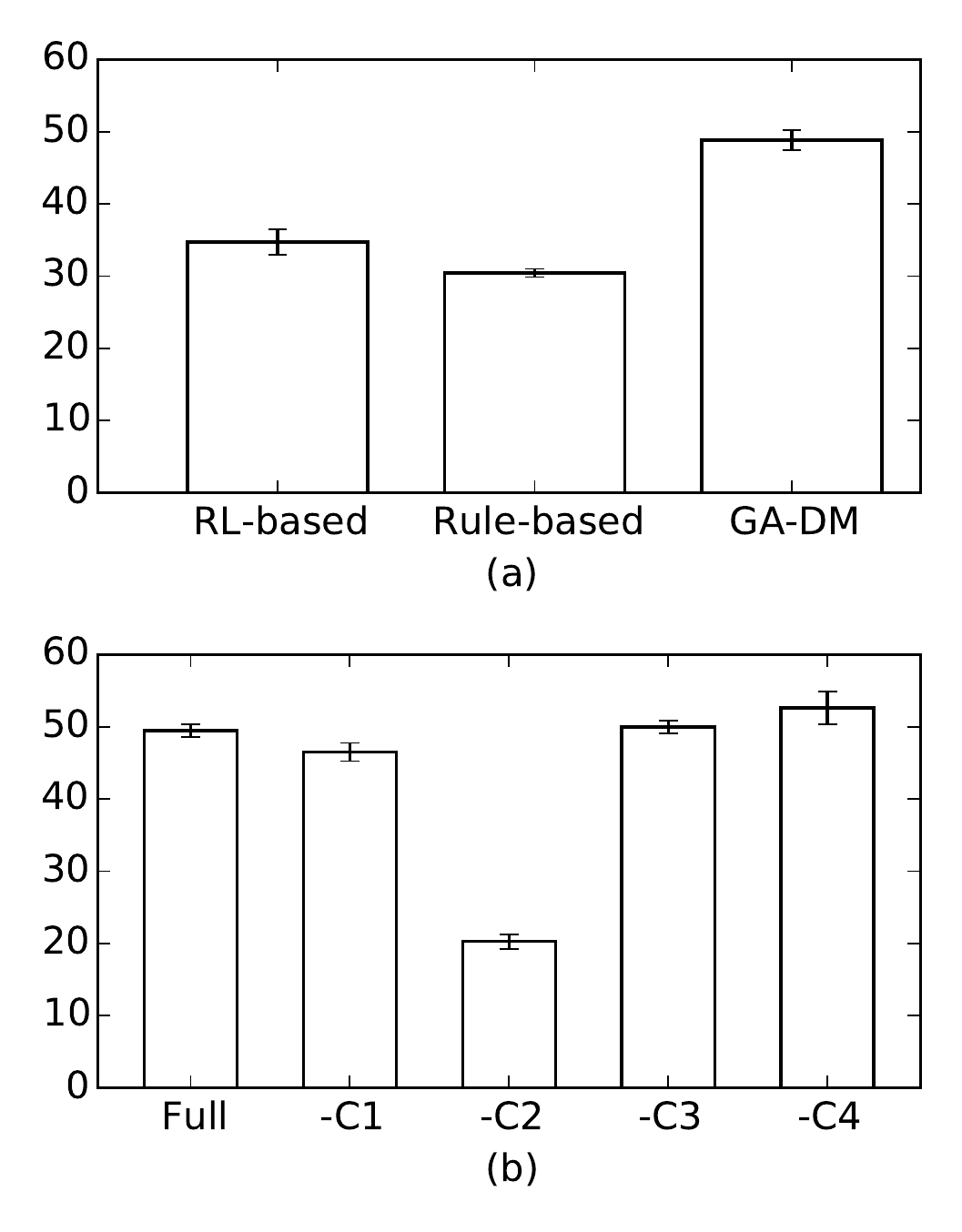}}
\caption{Overall testing performance of DM policies. 
(a) Performance of RL-based, rule-based and GA-DM policy.
(b) Performance of GA-DM with one clause disabled, where '-C1' means that C1 is disabled.
}
\label{res_testing}
\end{figure}

\begin{figure}[t]
\centering
\includegraphics[scale=0.65]{{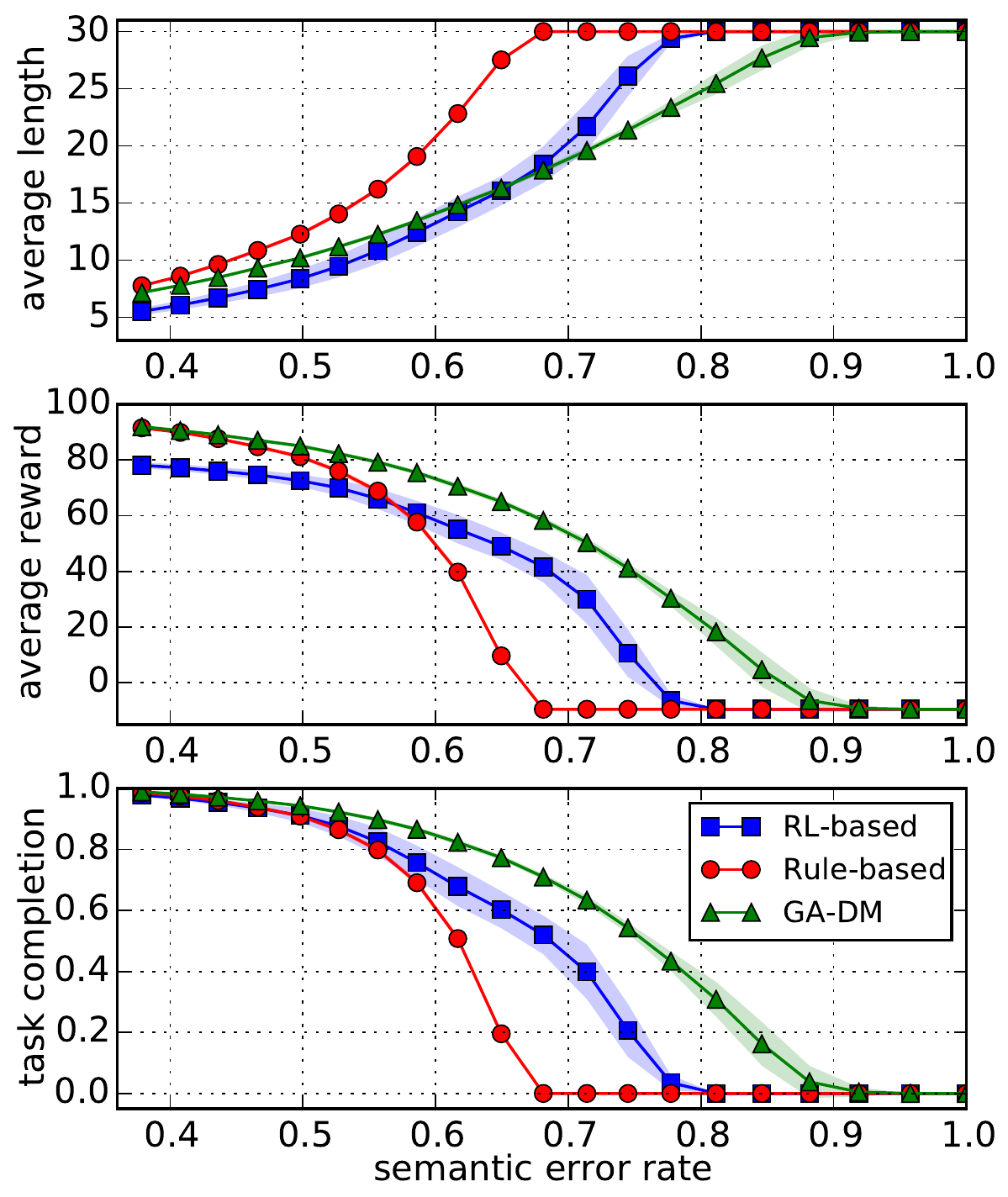}}
\caption{On-line evaluation of GA-DM using a simulated user
with fixed noise level.
Average cumulative reward, dialog length and task completion rate are plotted
against the error rate of the top SLU hypothesis,
which is used as a metric for environmental noise.
}
\label{res_simulation}
\end{figure}

\begin{figure}[t]
\centering
\includegraphics[scale=0.62]{{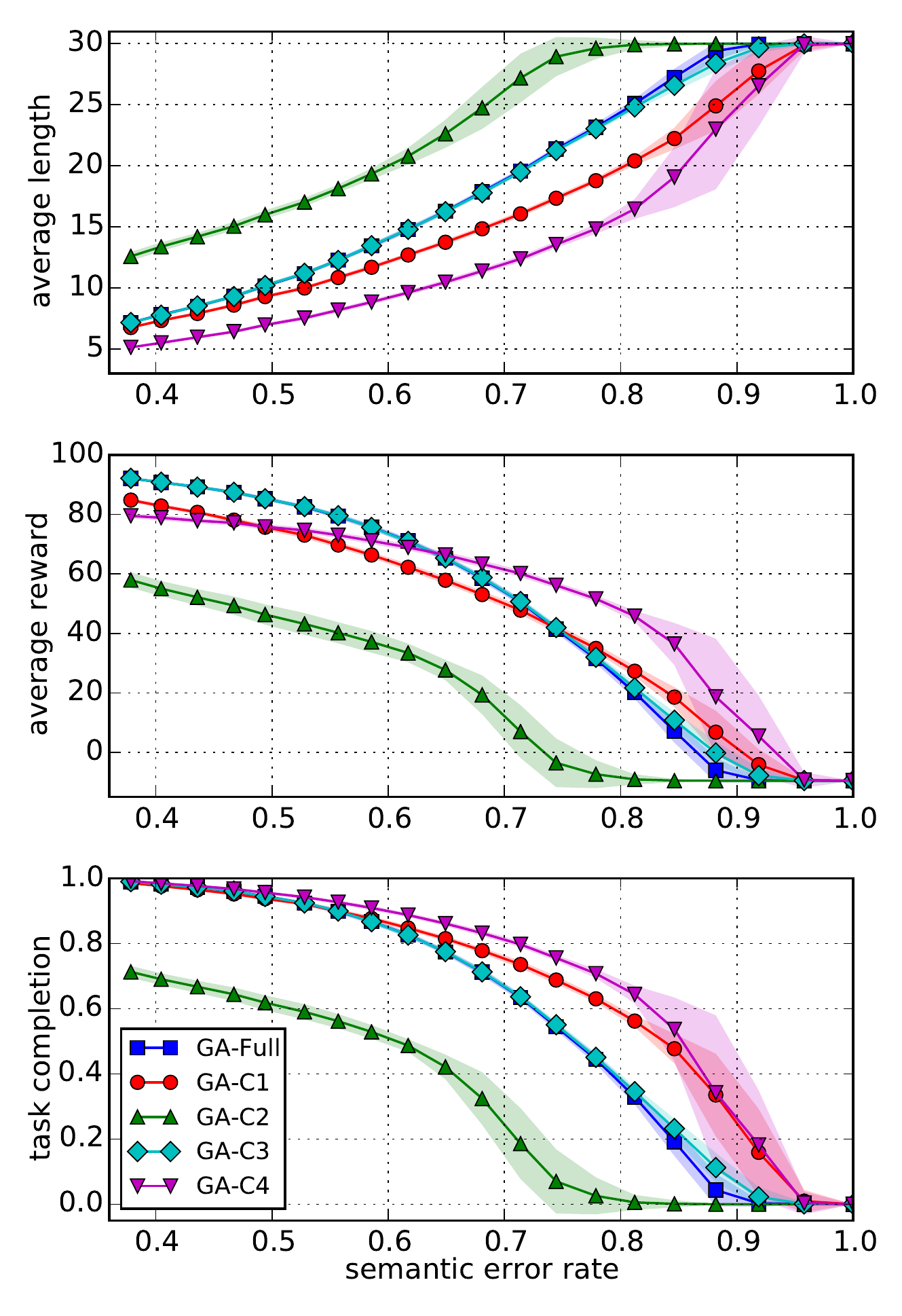}}
\caption{Performance of the model operating with fixed noise level when the major clauses c1-c4 are sequentially disabled
and retrained. Full GA-DM is the original model with all clauses enabled.}
\label{res_clauses}
\end{figure}

\begin{figure}[t]
\centering
\includegraphics[scale=0.65]{{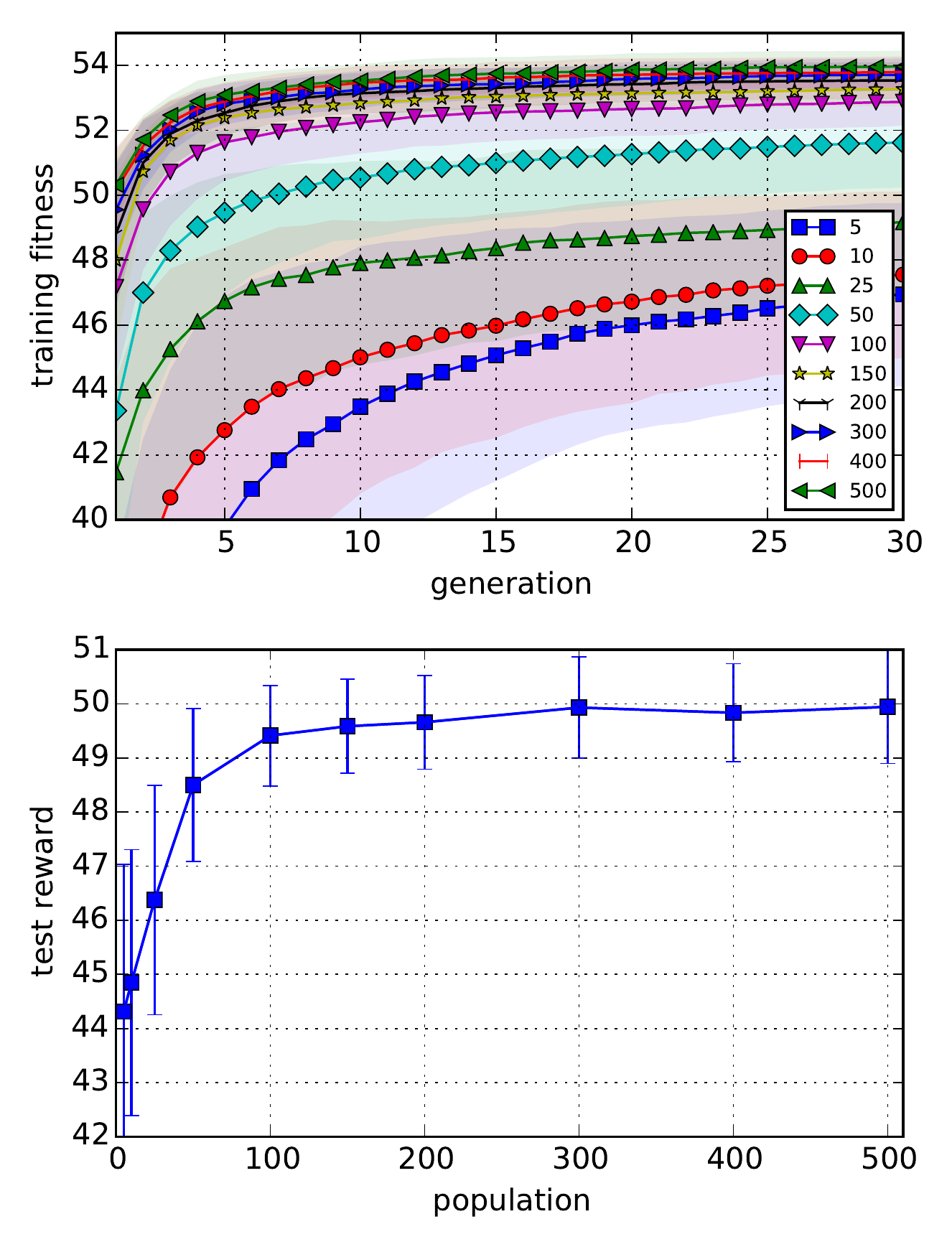}}
\caption{
Training fitness and testing performance of GA-DM 
when trained using different population size.
Standard deviations are plotted as shaded areas and error bars.
}
\label{res_varying_pop}
\end{figure}

During testing 1000 dialog sessions are conducted and the noise level is adjusted in the same way as in training.
We report both the overall performance (average reward received under a series of noise conditions) and 
the performance with a fixed noise level.
The overall testing performance of each DM is shown in Fig.\ref{res_testing}.
Performance when operating under fixed noise condition is shown in Fig.\ref{res_simulation} and \ref{res_clauses}.
The level of environmental noise is measured using
the semantic error rate of the top hypothesis of SLU results.
It should be emphasized that the noise levels shown in the results are the same ones used in training.
In addition to the GA-DM using the complete policy template,
the utility of each individual clause in the template is evaluated.
The four major clauses c1-c4 are disabled sequentially.
The resulted DMs are evaluated using the same settings,
and the testing results are shown along with the full-fledged GA-DM.
The effects of different GA population size are explored and reported in Fig.\ref{res_varying_pop}.

Since the DMs are optimized against the average reward received under several noise conditions,
the overall testing reward shown in Fig.\ref{res_testing} should be taken as the direct metric of performance.
The RL-based policy showed better overall performance than the rule-based one,
while GA-DM significantly outperformed both the rule-based and RL-based policies.
From Fig.\ref{res_simulation},
it can be seen that when the noise is low,
the rule-based DM is very competitive and
shows even better performance than the RL-based DM.
But when the noise level of the environment increases,
its performance degrades seriously,
while the RL-based DM is much more robust.
However, after tuning of the free parameters using GA,
the GA-DM outperforms both the other DMs on nearly all noise conditions.
Note the maximum noise level at which each DM could
successfully complete a dialog,
suggesting that the GA-DM is able to operate
under more adverse environment.
It is worth mentioning again that the rule-based DM
and GA-DM are instantiations of the same policy template.
The simulation results justify GA as an effective method
for DM policy optimization and
reveal the performance potential of
simple and yet human-interpretable DM policies.

It is interesting to make a comparison between RL and GA policy learning.
In DM policy optimization, the state space is often continuous and infinite.
In conventional RL, a model of the underlying optimal value function of the environment has to be designated.
The ability of the model to approximate the optimal value function 
is a key factor affecting the performance of the learnt policy.
However, the design of the model is often non-intuitive and complicated
since it operates in the value function space.
Expert knowledge is often difficult to be directly applied.
This fact can help to explain that in our experiments, 
the RL-based DM is not as competitive as the others when the noise level is low.
Since the noise level is varied during training,
the resulted learning environment is much more difficult to deal with than one with fixed noise condition.
Thus the linear model used is unlikely to perfectly match the underlying optimal value function
and cannot accommodate all types of condition.
In our experiment the RL policy has learnt to make a trade-off and adapted to conditions with high environmental noise for a better overall performance.
GA-DM tackles the problem from a different perspective.
It operates in policy space directly
and is much easier to incorporate expert knowledge.
In GA-DM a policy model is developed instead.
Equivalent assumptions about policy structure are often difficult to made in value function space.
Thus the resulted policy model can be more powerful and expressive than one for value function.

The relative utility of each clause of the policy template on the performance
is another interesting aspect to be investigated.
According to the results shown in
Fig.\ref{res_testing} (b) and Fig.\ref{res_clauses},
it can be observed that when C2 is disabled the performance drops seriously.
But to our surprise, when C4 is disabled,
the performance significantly boosts especially in high-noise regions.
The results show the relative utility of
each clause in the template
and reveal the necessity to optimize the structure of policy template.
This kind of structural optimization problem can also be solved using GA,
and we plan to study this kind of optimization in future work.

In GA the population size often influences the optimization efficiency.
The training fitness and testing performance using different population size is shown in Fig.\ref{res_varying_pop}.
We can observe that with an increasing population size,
the training and testing performance nearly monotonically increases.
This performance improvements are more obvious 
when the size is less than 100,
and are not noticeable above 300.
Because the elitism technique is used and
the fitness of the elitist individual is cached,
the training fitness improves steadily during training.

\subsection{On-corpus learning experiment}
\label{on_corpus_res}

The DSTC2 testing corpus is used for on-corpus DM learning and evaluation \cite{henderson_second_2014},
which is produced by a RL-based DM and consists of 1117 dialog sessions.
The full annotations of the dataset are released after 
the conclusion of the DSTC2 challenge.
The dialog state is the same as defined in the challenge,
and we use the results produced by the `focus' tracker
using the scripts provided by the DSTC2 organizer.
The dialog template used by GA comprises 9 condition-action clauses
and 6 free parameters.
The original corpus is equally split for
training and testing.

The reward function is defined as follows.
At each dialog turn the agent receives -10.0 reward.
If correct restaurants are offered to users, 100.0 points are rewarded.
But if the information is duplicate to that previously offered, 
-50.0 points are given.
If the restaurants offered do not meet user's demand,\\
-100.0 points are given.
The reward discounting rate $\gamma$ is set to 0.9.

In addition to the GA-based DMs trained using QPoints-regression
described in section \ref{on_corpus}, 
results of 3 additional DMs are shown for comparison.
\begin{enumerate}
\setlength\itemsep{0em}
\item \texttt{SL-Original} DM which is learned in a supervised way with the original dialog actions as training targets using the ExtraTrees classifier, represented as $\hat{P}(a|s)$.
\item \texttt{SL-MaxQ} supervised DM using the actions with maximum Q-value predicted by $\hat{Q}(s,a)$ as the supervised targets.
\item \texttt{ThresholdedQ} DM as described in \cite{lee_extrinsic_2014}, 
which selects the action with the maximum Q-value
predicted by $\hat{Q}(s, a)$
from the set of actions whose probabilities produced by 
$\hat{P}(a|s)$ are greater than $\delta$.
The threshold is used to constrain the behavior of RL policy,
in case of insufficient exploration.
\end{enumerate}

To make full use of the available data
and get a more stable estimation of the performance,
we conducted 12 re-sampling experiments
similar to the bootstrapping method, 
but avoid to use duplicate samples.
In each sub-experiment,
the dataset is reshuffled and split to get new
training and testing instances.
The average and standard deviation of
the results are shown in Table \ref{res_on_corpus}.

The \texttt{SL-MaxQ} DM which acts greedily upon $\hat{Q}(s, a)$ has poor performance on the test set
while being overrated on the training set,
probably as a result of insufficient exploration.
The \texttt{ThresholdedQ} DM
mitigates the problem to a great degree
by setting a simple threshold.
That heuristic is shared with the \texttt{QVal} fitness function.
\texttt{GA-QVal} outperforms all the other DMs
and is very stable across the re-sampling experiments
considering the relatively small standard deviation,
while the behavior \texttt{GA-NPoints} which is less consistent
results in an overall inferior performance.
Although \texttt{GA-QVal} is trained
under the guidance of an reinforcement learner $\hat{Q}(s, a)$,
its performance is superior to both
\texttt{SL-MaxQ} and \texttt{ThresholdedQ},
which should be attributed to the prior domain knowledge incorporated
into the policy template.
The DMs in bold outperform
\texttt{SL-Original} built by imitating the policy used in producing the corpus,
indicating the possibility of building a better and
yet human-comprehensible DM policy using a dialog corpus.

\section{Related work}
The subject of automatically optimizing dialog policies is a hot topic,
and many data-driven methods have been proposed among which
RL-based ones are the most popular.
There is some previous work on constraining the behavior of RL-based DM.
In \cite{williams_best_2008} Williams proposed to construct a hand-crafted DM 
and it produces a set of candidate actions for given dialog state,
from which the best one is chosen by a POMDP-DM.
Lison \cite{lison_hybrid_2015} proposed to use `probabilistic rule'
in specifying the transition and reward sub-models of the POMDP model.
The probabilistic rules are human-readable and less parameterized than conventional probability distribution,
thus reducing the free parameters of the POMDP model
and allowing the system designers to 
make use of domain knowledge in designing DM.
Our work bears some resemblance to \cite{lison_hybrid_2015}.
But we used the dialog policy template to specify a policy directly
and utilized GA to train the free parameters.

Henderson et al. proposed a hybrid learning method in \cite{henderson_hybrid_2008} 
to learn a policy on an existing dialog corpus by combining the results of supervised and reinforcement learning.
Pure RL on fixed dataset often shows irregular behavior due to the insufficient exploration problem.
Supervised learning (SL) is used to mitigate the problem and
the hybrid method shows better performance than pure SL or RL.
In this regard the QVal fitness function is similar in spirit
and the use of policy template can further constrain the DM behavior,
thus is suitable for off-line on-corpus learning.

One notable advantage of the GA-based DM over RL-based models is that 
the action structure can be changed during learning (only in on-line learning)
as described in section \ref{ga_dpt}.
While in RL, each action $a_i \in A$ must be invariant
otherwise the value function learned will be meaningless.
This characteristic is suitable for SDS engineering
since it can be difficult to determine the exact semantics
of a dialog action beforehand.
Further studies are needed in this regard.

\begin{table}[!t]
\begin{center}
\begin{tabular}{|r|r|r|}
\hline
DM  		&  Training		&  Testing \\\hline
GA-NPoints	& 98.46 (38.30)	&	89.52 (41.30) \\
GA-QVal		& 127.38 (5.59)	& \bf{129.29} (7.90)	   \\
SL-Original	& 115.63 (4.08) &	114.39 (6.07)	   \\
SL-MaxQ		& 245.19 (12.59)&	53.46 (36.06) \\
ThresholdedQ& 142.48 (4.22) & \bf{122.21} (4.36) \\\hline
\end{tabular}
\end{center}
\caption{\label{res_on_corpus} 
Estimated cumulative reward of DM policies 
on training and testing set.
Numbers in brackets are standard deviation
estimated by re-sampling experiments.
Only starting turns of a dialog are considered as described in section \ref{on_corpus_eval}.
\texttt{GA-NPoints} and \texttt{GA-QVal} are DMs trained using 
GA with $\NPoints$ and $\QVal$ fitness functions respectively.
}
\end{table}

\section{Conclusions and future work}
\label{conclusion}
In this paper we described a framework to train human-interpretable spoken
dialog management policies using genetic algorithm.
Two kinds of fitness functions were used,
i.e., one based on interacting with a simulated user
and the other on a dialog corpus which is more sample-efficient.
We set up an online simulation environment
and used the DSTC2 corpus for off-line on-corpus training and evaluation.
The results show that by using domain language
and setting appropriate free parameters,
the performance of simple rule-based DM policies can be largely improved,
and can even outperforms those trained using reinforcement learning.
According to our knowledge,
this is the first time that genetic algorithm is applied to DM optimization.
Another advantage is its ability to optimize the structure of system actions.
This framework is very suitable to upgrade existing SDSs using rule-based DM,
by using collected data to optimize
the newly specified free parameters.


This research is still preliminary and
several aspects need further investigation,
especially the effects of fitness functions.
The search space of dialog policy in GA can be expanded
by allowing the condition-action expressions to be reordered
and partially disabled.
The structural learning in system actions also needs further studies.
We hope this work can help
to build better and practical
spoken dialog systems.
\bibliographystyle{IEEEbib}
\bibliography{gadm_ref}

\begin{thebibliography}{10}

\bibitem{williams_partially_2007}
Jason~D. Williams and Steve Young,
\newblock ``Partially observable {Markov} decision processes for spoken dialog
  systems,''
\newblock {\em Computer Speech \& Language}, vol. 21, no. 2, pp. 393--422,
  2007.

\bibitem{young_hidden_2010}
Steve Young, Milica Gašić, Simon Keizer, François Mairesse, Jost Schatzmann,
  Blaise Thomson, and Kai Yu,
\newblock ``The {Hidden} {Information} {State} model: {A} practical framework
  for {POMDP}-based spoken dialogue management,''
\newblock {\em Computer Speech \& Language}, vol. 24, no. 2, pp. 150--174,
  2010.

\bibitem{lee_example-based_2009}
Cheongjae Lee, Sangkeun Jung, Seokhwan Kim, and Gary~Geunbae Lee,
\newblock ``Example-based dialog modeling for practical multi-domain dialog
  system,''
\newblock {\em Speech Communication}, vol. 51, no. 5, pp. 466--484, 2009.

\bibitem{lison_structured_2014}
Pierre. Lison,
\newblock {\em Structured {Probabilistic} {Modelling} for {Dialogue}
  {Management}},
\newblock Ph.D. thesis, University of Oslo, 2014.

\bibitem{young_pomdp-based_2013}
Steve Young, Milica Gašić, Blaise. Thomson, and Jason~D. Williams,
\newblock ``{POMDP}-{Based} {Statistical} {Spoken} {Dialog} {Systems}: {A}
  {Review},''
\newblock {\em Proceedings of the IEEE}, vol. 101, no. 5, pp. 1160--1179, 2013.

\bibitem{paek_reinforcement_2006}
Tim Paek,
\newblock ``Reinforcement {Learning} for {Spoken} {Dialogue} {Systems}:
  {Comparing} {Strengths} and {Weaknesses} for {Practical} {Deployment},''
\newblock Tech. {R}ep. MSR-TR-2006-62, Microsoft Research, 2006.

\bibitem{engel_reinforcement_2005}
Yaakov Engel, Shie Mannor, and Ron Meir,
\newblock ``Reinforcement learning with {Gaussian} processes,''
\newblock in {\em Proceedings of the 22nd international conference on {Machine}
  learning}. 2005, pp. 201--208, ACM.

\bibitem{gasic_gaussian_2010}
Milica Gašić, Filip Jurčíček, Simon Keizer, François Mairesse, Blaise
  Thomson, Thomson, Kai Yu, and Steve Young,
\newblock ``Gaussian {Processes} for {Fast} {Policy} {Optimisation} of
  {POMDP}-based {Dialogue} {Managers},''
\newblock in {\em Proceedings of the 11th {Annual} {Meeting} of the {Special}
  {Interest} {Group} on {Discourse} and {Dialogue}}, Stroudsburg, PA, USA,
  2010, {SIGDIAL} '10, pp. 201--204, Association for Computational Linguistics.

\bibitem{whitley_genetic_1994}
Darrell Whitley,
\newblock ``A genetic algorithm tutorial,''
\newblock {\em Statistics and Computing}, vol. 4, no. 2, pp. 65--85, June 1994.

\bibitem{schatzmann_agenda-based_2007}
Jost Schatzmann, Blaise Thomson, Karl Weilhammer, Hui Ye, and Steve Young,
\newblock ``Agenda-based user simulation for bootstrapping a {POMDP} dialogue
  system,''
\newblock in {\em Human {Language} {Technologies} 2007: {The} {Conference} of
  the {North} {American} {Chapter} of the {Association} for {Computational}
  {Linguistics}; {Companion} {Volume}, {Short} {Papers}}. 2007, pp. 149--152,
  Association for Computational Linguistics.

\bibitem{henderson_hybrid_2008}
James Henderson, Oliver Lemon, and Kallirroi Georgila,
\newblock ``Hybrid {Reinforcement}/{Supervised} {Learning} of {Dialogue}
  {Policies} from {Fixed} {Data} {Sets},''
\newblock {\em Computational Linguistics}, vol. 34, no. 4, pp. 487--511, 2008.

\bibitem{Pietquin2011}
Olivier Pietquin, Matthieu Geist, Senthilkumar Chandramohan, and Herv\'{e}
  Frezza-Buet,
\newblock ``{Sample-efficient batch reinforcement learning for dialogue
  management optimization},''
\newblock {\em ACM Transactions on Speech and Language Processing (TSLP)}, vol.
  7, no. 3, pp. 7, 2011.

\bibitem{ernst_tree-based_2005}
Damien Ernst, Pierre Geurts, and Louis Wehenkel,
\newblock ``Tree-based batch mode reinforcement learning,''
\newblock in {\em Journal of {Machine} {Learning} {Research}}, 2005, pp.
  503--556.

\bibitem{geurts_extremely_2006}
Pierre Geurts, Damien Ernst, and Louis Wehenkel,
\newblock ``Extremely randomized trees,''
\newblock {\em Machine Learning}, vol. 63, no. 1, pp. 3--42, Mar. 2006.

\bibitem{lee_extrinsic_2014}
Sungjin Lee,
\newblock ``Extrinsic {Evaluation} of {Dialog} {State} {Tracking} and
  {Predictive} {Metrics} for {Dialog} {Policy} {Optimization},''
\newblock in {\em 15th {Annual} {Meeting} of the {Special} {Interest} {Group}
  on {Discourse} and {Dialogue}}, 2014, p. 310.

\bibitem{henderson_second_2014}
Matthew Henderson, Blaise Thomson, and Jason~D. Williams,
\newblock ``The second dialog state tracking challenge,''
\newblock in {\em 15th {Annual} {Meeting} of the {Special} {Interest} {Group}
  on {Discourse} and {Dialogue}}, 2014, p. 263.

\bibitem{williams_best_2008}
Jason~D. Williams,
\newblock ``The best of both worlds: unifying conventional dialog systems and
  {POMDPs}.,''
\newblock in {\em {INTERSPEECH}}, 2008, pp. 1173--1176.

\bibitem{lison_hybrid_2015}
Pierre Lison,
\newblock ``A hybrid approach to dialogue management based on probabilistic
  rules,''
\newblock {\em Computer Speech \& Language}, vol. 34, no. 1, pp. 232--255,
  2015.

\end{thebibliography}

\end{document}